\documentclass[10pt,prd,aps,twocolumn,preprintnumbers,showpacs, nofootinbib,superscriptaddress,notitlepage]{revtex4-2}

\usepackage{slashed}
\usepackage{graphicx,color}
\usepackage{epsfig}
\usepackage{subfigure}
\usepackage{epsfig}
\usepackage{multirow}
\usepackage{epstopdf}
\usepackage{amsmath}
\usepackage{dcolumn}
\usepackage{bm}
\usepackage{color}
\usepackage{ulem}
\usepackage{enumitem}
\usepackage[colorlinks,
linkcolor=red,
anchorcolor=blue,
citecolor=blue
]{hyperref}

\begin{document}

\title{Revealing Neutrino Mass Ordering at CEPC and FCC-ee}

\author{Wei Liu}
\affiliation{ Department of Applied Physics and MIIT Key Laboratory of Semiconductor Microstructure and Quantum Sensing,
Nanjing University of Science and Technology, Nanjing 210094, China}

\author{Supriya Senapati}
\affiliation{ Department of Applied Physics and MIIT Key Laboratory of Semiconductor Microstructure and Quantum Sensing,
Nanjing University of Science and Technology, Nanjing 210094, China}

\author{Jin Sun}
\email{sunjin0810@ibs.re.kr(Contact author)}
\affiliation{Particle Theory and Cosmology Group, Center for Theoretical Physics of the Universe, Institute for Basic Science (IBS), Daejeon 34126, Korea }

\begin{abstract}
The ordering of neutrino masses remains a key unknown in particle physics and cosmology. While upcoming oscillation experiments are expected to determine the mass ordering at low energies, it is important to explore complementary probes that access the underlying mechanism of neutrino mass generation. In this work, we show that future high-energy electron–positron colliders can provide sensitivity to the neutrino mass ordering through the lepton-flavor structure of heavy neutral lepton (HNL) interactions. In the minimal Type-I seesaw scenario with two nearly degenerate HNLs, the flavor composition of the heavy–light neutrino mixings is strongly correlated with the light-neutrino mass spectrum, leading to distinct collider signatures for normal and inverted mass orderings. We demonstrate that future $Z$ factories such as CEPC and FCC-ee can probe these flavor patterns over a wide region of parameter space, establishing collider searches for HNLs as a complementary approach to neutrino mass ordering studies.

\end{abstract}

\maketitle
\preprint{$\begin{gathered}\includegraphics[width=0.04\textwidth]{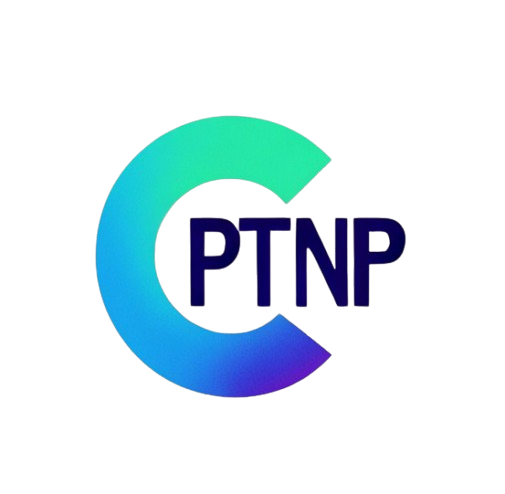}\end{gathered}$\,CPTNP-2026-002}
\baselineskip=15pt
\preprint{CTPU-PTC-26-02}

\noindent{\bf \textit{Introduction}} Neutrino oscillation experiments~\cite{PDG} allow two possible neutrino mass orderings: the \textbf{normal hierarchy (NH)}, $m_3 > m_2 > m_1$, and the \textbf{inverted hierarchy (IH)}, $m_2 > m_1 > m_3$, with neutrino masses below the eV scale. Determining the true neutrino mass ordering is a central open question in particle physics, with profound implications for the origin of neutrino mass, neutrinoless double beta decay, and the matter--antimatter asymmetry of the Universe. While upcoming experiments such as JUNO~\cite{JUNO:2015sjr}, DUNE~\cite{DUNE:2020lwj}, and Hyper-Kamiokande~\cite{Hyper-Kamiokande:2018ofw} aim to resolve this question through precision measurements of neutrino oscillations 
at low energies, complementary probes from other frontiers are equally essential.

Heavy neutral leptons (HNLs), predicted in the Type-I seesaw mechanism~\cite{Minkowski:1977sc,Yanagida:1980xy,Gell-Mann:1979vob,Mohapatra:1979ia}, provide a complementary probe of neutrino mass generation at high-energy colliders. These heavy Majorana fermions mix weakly with the active neutrinos, with mixings parametrized by $|U_{\alpha N}|^2$ ($\alpha=e,\mu,\tau$), and $N$ represents the mass eigenstates of the HNLs. However, existing collider searches mainly focused on single-flavor mixing scenarios, they are generally insensitive to  the neutrino mass ordering~\cite{Kaneta:2016vkq, Das:2017zjc, Das:2017rsu, Bondarenko:2018ptm, Bryman:2019bjg, Balaji:2019fxd, Liu:2021akf, Abdullahi:2022jlv, Zhang:2023nxy, Barducci:2023hzo, Liu:2023klu, Liu:2022kid,Han:2021pun, Das:2017nvm, Maiezza:2015lza, Deppisch:2018eth, Mason:2019okp, Accomando:2016rpc, Gago:2015vma, Jones-Perez:2019plk, Liu:2022ugx, Li:2023dbs, Deppisch:2023sga, Liu:2024fey, Wang:2024mrc, Wang:2024prt}. In this Letter, we show that collider experiments can determine the mass ordering   by exploiting the flavor structure of HNL interactions.

To reproduce neutrino oscillation data, at least two HNLs are required. In the minimal seesaw scenario with two nearly degenerate HNLs, the HNL flavor mixings are strongly correlated with the light-neutrino mass spectrum. In particular, the mixings inherit the flavor structure of the heaviest light-neutrino states, leading to a suppressed electron mixing in the NH,
$|U_{eN}^2| \ll |U_{\mu N}^2| \sim |U_{\tau N}^2|$,
whereas comparable mixings among all flavors are expected in the IH,
$|U_{eN}^2| \sim |U_{\mu N}^2| \sim |U_{\tau N}^2|$~\cite{Drewes:2016jae, Tastet:2021vwp,Drewes:2022akb, Abada:2022wvh,Drewes:2024bla}.
Consequently, the comparison of lepton-flavor final states in HNL decays provides a indirect handle on the neutrino mass ordering.

Future $Z$ factories, such as CEPC and FCC-ee, offer an ideal environment to test this idea due to the large statistics of HNLs production. We quantify the sensitivity to the mass ordering in the seesaw parameter space $(m_N,\,U_{\mathrm{tot}}^2)$ and demonstrate that colliders can provide a complementary determination of the mass ordering.\\

\noindent{\bf \textit{Lepton Flavor Mixing of the HNL}}
In the minimal Type-I seesaw scenario with two nearly degenerate HNLs
$N_I$ $(I=1,2)$, the heavy--light mixing parameters $\Theta_{\alpha I}$ not only govern the production and decay of HNLs at colliders, but also encode the flavor structure imprinted by the light-neutrino mass spectrum. 
As a result, the relative strengths of the HNL couplings to the electron, muon, and tau flavors depend sensitively on the neutrino mass ordering.
In particular, the NH predicts a suppressed electron mixing due to the vanishing lightest neutrino mass, whereas the IH leads to a more democratic flavor pattern.
This makes the lepton-flavor composition of HNL interactions a sensitive probe of the mass ordering.

The experimentally measurable quantities $U_{\alpha N}^2$ arise from the cumulative effects of both $N_1$ and $N_2$, expressed as~\cite{Eijima:2018qke}
\begin{eqnarray}\label{eq:MO}
&&  U_{\alpha N}^2=\sum_I |\Theta_{\alpha I}|^2=\frac{1}{2m_N}\left(|c_\alpha^+|^2 x_\omega^2+|c_\alpha^-|^2 x_\omega^{-2}\right),\nonumber\\
&&\mbox{NH}:\quad  c_\alpha^\pm=i V_{\alpha2}^{\rm PMNS}\sqrt{m_2}\pm V_{\alpha 3}^{\rm PMNS}\sqrt{m_3}\;,\nonumber\\
 && \qquad\quad  U_{\rm tot}^2=\sum_{\alpha, I} |\Theta_{\alpha I}|^2= \frac{(m_2+m_3)}{2m_N}(x_\omega^2+x_{\omega}^{-2})\;,\nonumber\\
&&\mbox{IH}:\quad  c_\alpha^\pm=i V_{\alpha1}^{\rm PMNS}\sqrt{m_1}\pm V_{\alpha 2}^{\rm PMNS}\sqrt{m_2}\;,\nonumber\\
 && \qquad\quad  U_{\rm tot}^2=\sum_{\alpha, I} |\Theta_{\alpha I}|^2= \frac{(m_1+m_2)}{2m_N}(x_\omega^2+x_\omega^{-2})\;.
\end{eqnarray}
Here $U_{\alpha N}^2$ and $U_{\rm tot}^2$  quantify the HNL mixing to a particular flavor and three flavors in total, respectively. And the rest of quantities are defined in the appendix.
Eq.~(\ref{eq:MO}) shows explicitly that, in the minimal seesaw with two HNLs, 
the flavor composition of the heavy--light mixing is not arbitrary but is fully determined 
by the light-neutrino mass spectrum and the PMNS mixing matrix. 
Consequently, the HNL flavor ratios provide a direct imprint of the mass ordering.

\begin{figure}[!t]
\centering
{\includegraphics[width=.99\columnwidth]{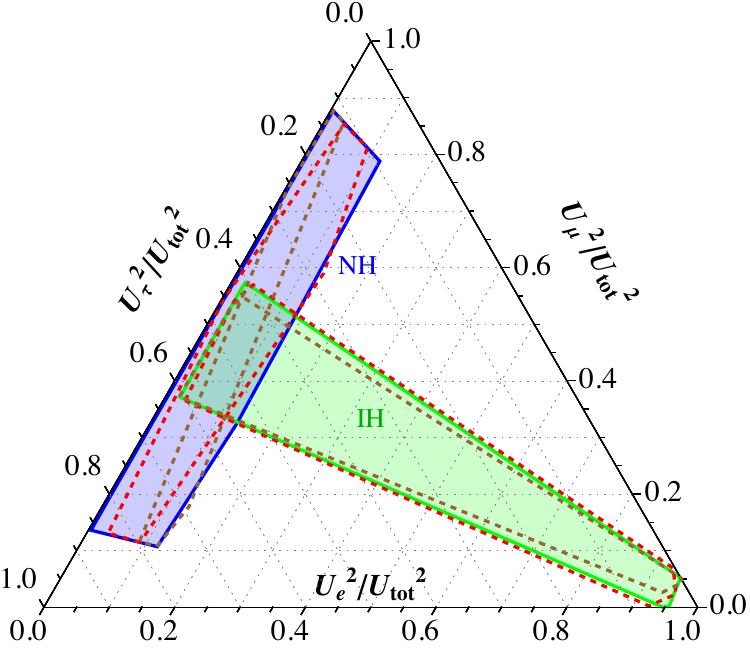}}
	\caption{ Ternary plot showing the 
    $3\sigma$ allowed region of flavor-mixing  ratios $U^2_\alpha/U^2_{tot}$, $\alpha=e,\mu,\tau$
    for  NH (blue) and IH (green).
     They are obtained by  NuFIT 6.0~\cite{Esteban:2024eli,Nufit} 
     and JUNO's new result~\cite{JUNO:2025gmd}.
     The dashed red (brown) curve shows the  region under the best-fit CP phase from T2K (NO$\nu$A), assuming a future uncertainty of $8^\circ$ ($7^\circ$) from DUNE+$\mu$THEIA~\cite{Ge:2022iac}.
	}
	\label{fig:CP}
\end{figure}

Following Eq.~(\ref{eq:MO}), we scan the neutrino oscillation parameters within their 
$3\sigma$ allowed ranges from NuFIT~\cite{Esteban:2024eli,Nufit} and JUNO’s latest result~\cite{JUNO:2025gmd} 
to obtain the ternary distributions shown in Fig.~\ref{fig:CP}. 
In the NH case, the absence of the $\nu_1$ contribution, together with 
$|V_{e1}^{\rm PMNS}|\gg|V_{e2,3}^{\rm PMNS}|$, leads to a significantly suppressed electron mixing of the HNL. 
As a result, the allowed range of $U_{eN}^2/U_{\rm tot}^2$ is confined to 
$(3.1\times10^{-3},\,0.14)$, as summarized in Table~\ref{tab:range}.

Improved precision in neutrino oscillation measurements can further reduce the allowed parameter space. 
Although JUNO’s first results already demonstrate this potential~\cite{JUNO:2025gmd}, 
sizable uncertainties remain. 
Among the oscillation parameters, the Dirac CP phase $\delta_D$ is currently the least constrained 
and exhibits a notable tension between T2K~\cite{T2K:2023smv} and NO$\nu$A~\cite{NOvA:2021nfi}, 
with best-fit values of $\delta_D\simeq-\pi/2$ (T2K) and $\delta_D\simeq148^\circ$ (NO$\nu$A)\footnote{
Recent combined analyses have partially reduced this tension~\cite{T2K:2025wet}.}. 
To illustrate its impact, Fig.~\ref{fig:CP} shows the predicted NH regions corresponding to the 
best-fit $\delta_D$ values from T2K (red dashed) and NO$\nu$A (brown dashed), 
assuming future uncertainties of $8^\circ$ and $7^\circ$, respectively, 
achievable with DUNE and $\mu$THEIA~\cite{Ge:2022iac}. 
In contrast, the IH region remains largely unchanged. 
This demonstrates that future precision measurements of $\delta_D$ will significantly sharpen 
the HNL flavor predictions.\\

\begin{table}[t]
    \centering
    \small
    \begin{tabular}{|c|c|c|c|}
        \hline
        Mass Ordering & $U_{eN}^2/U_{\rm tot}^2$ & $U_{\mu N}^2/U_{\rm tot}^2$ 
        & $U_{\tau N}^2/U_{\rm tot}^2$ \\
        \hline
        NH & ($3.1 \times 10^{-3}$, 0.14) & (0.11, 0.88) &  (0.09, 0.86) \\
        \hline
        IH & (0.021, 0.96) & (0, 0.57) & (0, 0.61)\\
        \hline
    \end{tabular}
    \caption{The range of $U_{\alpha N}^2/U_{\rm tot}^2$ for NH/IH mass orderings.}
    \label{tab:range}
\end{table}

\noindent{\bf \textit{Sensitivity on neutrino mass ordering}}
The distinct flavor structures of HNL mixings associated with NH and IH 
provide a novel way to probe the neutrino mass ordering through the lepton-flavor composition 
of HNL decays at colliders, as discussed below.

To achieve large HNL statistics, we focus on future high-luminosity lepton colliders, 
in particular the Circular Electron--Positron Collider (CEPC)~\cite{CEPCStudyGroup:2018rmc} 
and the Future Circular Collider in the electron--positron mode (FCC-ee)~\cite{FCC:2018evy}. 
Our analysis targets the $Z$ pole, corresponding to a center-of-mass energy of 
$\sqrt{s}=91.2~\mathrm{GeV}$, with an integrated luminosity of 
$\mathcal{L}\simeq200~\mathrm{ab}^{-1}$.

The signal process consists of single-HNL production in association with a neutrino or antineutrino, 
followed by its decay predominantly into a charged lepton and a pair of jets,
\begin{equation}
e^+ e^- \to N \nu~(\bar{\nu}), \qquad N \to \ell_\alpha j j \, ,
\end{equation}
where $\ell_\alpha = e,\mu$ denotes the charged lepton in the final state. 
Final states involving $\tau$ leptons are neglected due to the limited reconstruction efficiency and larger associated uncertainties.
To ensure precise lepton-flavor identification and sufficient event statistics, 
we focus on prompt HNL decays, requiring the decay length in the laboratory frame to satisfy 
$L_N \lesssim 1~\mathrm{mm}$. 
Alternative facilities, such as a muon collider~\cite{Kwok:2023dck, Li:2023tbx, Mekala:2023diu} 
and $\mu$TRISTAN~\cite{Das:2024kyk}, can also probe HNL production. 
However, for the processes considered here, the corresponding production cross sections 
are significantly suppressed compared to those at CEPC and FCC-ee. 
We therefore restrict our collider analysis to electron--positron machines.

Using \textsc{MG5\_aMC@NLO}~\cite{Alwall:2014hca}, event samples are generated with selection criteria requiring 
leptons and jets to lie within a pseudorapidity range of $|\eta|<2.5$ and to satisfy 
$p_T(j/\ell)>10~\mathrm{GeV}$. 
The expected number of signal events in each lepton flavor channel is given by
\begin{equation}
N_{S,\alpha}
=\sigma(Z)\times \mathrm{Br}(Z\to N\nu~(\bar{\nu}))
\times \mathrm{Br}(N\to \ell_\alpha j j)
\times \mathcal{L}\times \epsilon ,
\end{equation}
where $\mathrm{Br}(Z\to N\nu~(\bar{\nu}))\propto U_{\rm tot}^2$ and 
$\mathrm{Br}(N\to \ell_\alpha j j)\propto U_{\alpha N}^2/U_{\rm tot}^2$. 
The overall efficiency is factorized as $\epsilon\simeq\epsilon_{\rm prompt}\times\epsilon_{\rm recon}$.
The prompt-decay efficiency,
$\epsilon_{\rm prompt}\simeq1-e^{-l_{\rm dec}/L_N}$,
accounts for HNL decays occurring within the primary vertex resolution 
$l_{\rm dec}\simeq1~\mathrm{mm}$, where the decay length $L_N$ is given in the appendix. 
For electrons and muons, both the CEPC and FCC-ee detectors are designed to achieve 
near-perfect reconstruction and identification performance. 
Full detector simulations indicate reconstruction efficiencies exceeding $99\%$ 
for isolated leptons over a wide angular acceptance~\cite{IDEAStudyGroup:2025gbt,Ai:2024nmn}.

Hence, for short-lived HNLs, the number of signal events can be well approximated as
\begin{equation}
N_{S,\alpha} \simeq \mathcal{O}(10^{11}) \times U_{\rm tot}^2 \times R_\alpha
\equiv N_0 \times R_\alpha ,
\end{equation}
where $R_\alpha \equiv U_{\alpha N}^2/U_{\rm tot}^2$ encodes the flavor composition of the active--sterile mixing, 
and $N_0 \sim \mathcal{O}(10^{11}) \times U_{\rm tot}^2$ denotes the total number of signal events, 
which is independent of flavor.

The dominant Standard Model background arises from irreducible diboson processes. 
Applying the same selection criteria, we find
\begin{equation}
N_{B,\alpha} \simeq 1.4 \times 10^5 ,
\end{equation}
per flavor for $\alpha=e,\mu$.

To probe the neutrino mass ordering, we analyze the signal and background distributions 
in the two-flavor plane $(R_e, R_\mu)$, which directly maps onto the ternary regions shown in Fig.~\ref{fig:CP}. 
For a fixed value of $U_{\rm tot}^2$, we define the detection significance as
\begin{equation}
\chi^2_{\rm NH/IH}
=
\sum_{\alpha=e,\mu}
\frac{
\left[
N_0\left(
R^{\rm obs}_\alpha - R^{\rm NH/IH}_\alpha
\right)
\right]^2
}{
N_0 R^{\rm NH/IH}_\alpha + N_{B,\alpha}
} ,
\end{equation}
where $R_\alpha^{\rm obs}$ denotes the observed flavor fraction, and 
$R_\alpha^{\rm NH/IH}$ is the theoretical prediction under the normal or inverted mass ordering. 
Requiring $\min(\chi^2_{\rm NH/IH})>4$, we can exclude the NH or IH hypothesis at approximately the $2\sigma$ level.

\begin{figure*}
    \centering
    \subfigure[\label{fig:31}]
 	{\includegraphics[width=.245\textwidth]{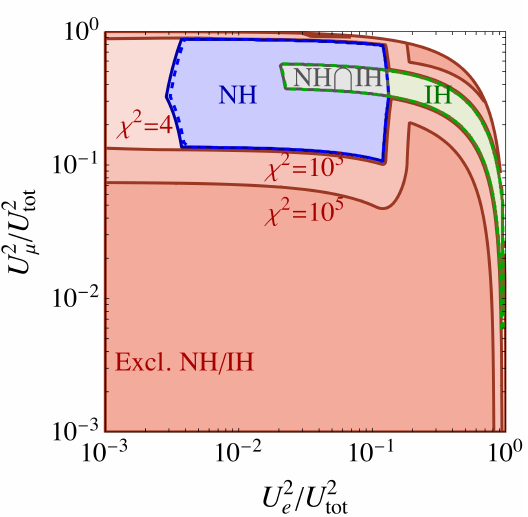}}
    \subfigure[\label{fig:32}]
 	{\includegraphics[width=.245\textwidth]{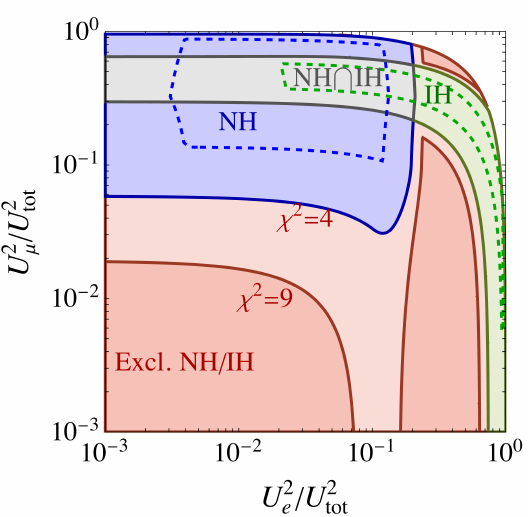}}
    \subfigure[\label{fig:33}]
 	{\includegraphics[width=.245\textwidth]{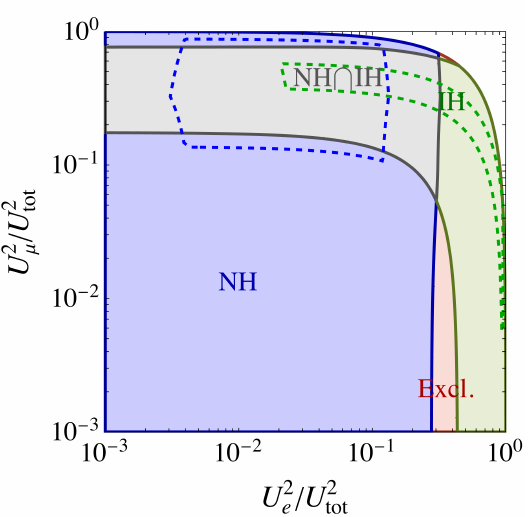}}
    \subfigure[\label{fig:34}]
    {\includegraphics[width=.245\textwidth]{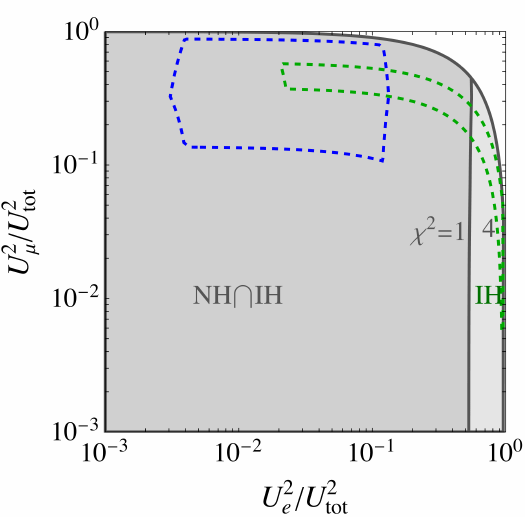}}
    
    \caption{Sensitivity of the neutrino mass orderings in the plane of $(U_{e N}^2/U_{\rm tot}^2, U_{\mu N}^2/U_{\rm tot}^2)$,
    at $\sqrt{s}=$ 91.2 GeV CEPC/FCC-ee with $\mathcal{L} \approx 200~ab^{-1}$, for $U^2_{\rm tot} \approx 10^{-5}$~(a), $10^{-7}$~(b), $10^{-8}$~(c), and $4 \times 10^{-9}$~(d), respectively, when $m_N \sim 50$~GeV.
    }
    \label{fig:bench}
\end{figure*}

In Fig.~\ref{fig:bench}, we present the sensitivity to the neutrino mass ordering 
in the $(R_e, R_\mu)$ plane for $U_{\rm tot}^2\simeq10^{-5}$ (a), $10^{-7}$ (b), 
$10^{-8}$ (c), and $4\times10^{-9}$ (d), respectively, assuming $m_N\simeq50~\mathrm{GeV}$. 
 $\chi_{NH/IH}^2$ is evaluated by taking $R_\alpha^{\rm obs}$ to coincide with the coordinates 
in the $(R_e, R_\mu)$ plane, while $R_\alpha^{\rm NH/IH}$ corresponds to the regions 
predicted for  NH (blue dashed) and IH (green dashed) hypotheses.

Following Table~\ref{tab:region}, the figure distinguishes four regions according to 
their capability to test the NH and IH hypotheses. 
As $U_{\rm tot}^2$ increases in Fig.~\ref{fig:bench}, the resulting growth in the $\chi_{NH/IH}^2$ 
leads to three correlated effects: 
(i) an expansion of the excluded region, 
(ii) a contraction of the ambiguous region where the NH and IH cannot be distinguished, 
and (iii) a convergence of the NH- or IH-favored regions toward their respective 
theoretically predicted parameter spaces. 
Taken together, these effects substantially enhance the discrimination power 
between the two neutrino mass orderings.

\begin{table}
    \centering
    \small
    \begin{tabular}{|c|c|c|c|}
        \hline
        Region & Color & $\min(\chi^2_{NH})$ 
        & $\min(\chi^2_{IH})$ \\
        \hline
        NH Favor & Blue & $< 4$ &  $> 4$ \\
        \hline
        IH Favor & Green & $> 4$ & $< 4$\\
        \hline
        NH $\cap$ IH Favor & Gray & $< 4$ & $< 4$\\
        \hline
        Excluded & Red & $> 4$& $> 4$\\
        \hline
    \end{tabular}
    \caption{The NH, IH, NH/IH Favor and Excluded Regions in Fig.~\ref{fig:bench} and their range of $\min(\chi_{NH/IH}^2)$.}
    \label{tab:region}
\end{table}

To quantify the overall discrimination power, we define the fraction of parameter space 
(in $\log_{10}$ scale) for which the neutrino mass ordering can be unambiguously identified,
\begin{equation}
K = 1 - \frac{S_{\rm NH\cap IH}}{S_{\rm tot}} \, ,
\end{equation}
where $S_{\rm NH\cap IH}$ denotes the area of the region in which both the NH and IH hypotheses 
are allowed, and $S_{\rm tot}$ is the total area spanned by the theoretically predicted 
NH and IH parameter spaces. 
The quantity $K$ therefore provides a global measure of the fraction of the allowed seesaw 
parameter space that permits an unambiguous determination of the neutrino mass ordering, 
rather than a confidence level associated with a single parameter point.

Owing to the intrinsic overlap between the predicted NH and IH regions, 
we find $K\lesssim0.90$ even in the limit of large event statistics. 
When $N_0 \lesssim 2\sqrt{\sum_{\alpha=e,\mu} N_{B,\alpha}}$, 
the excluded region disappears, while the overlap region ($\mathrm{NH}\cap\mathrm{IH}$) 
covers almost the entire parameter space, driving $K\simeq0$.

\begin{figure}[t!]
\centering
\includegraphics[width=\columnwidth]{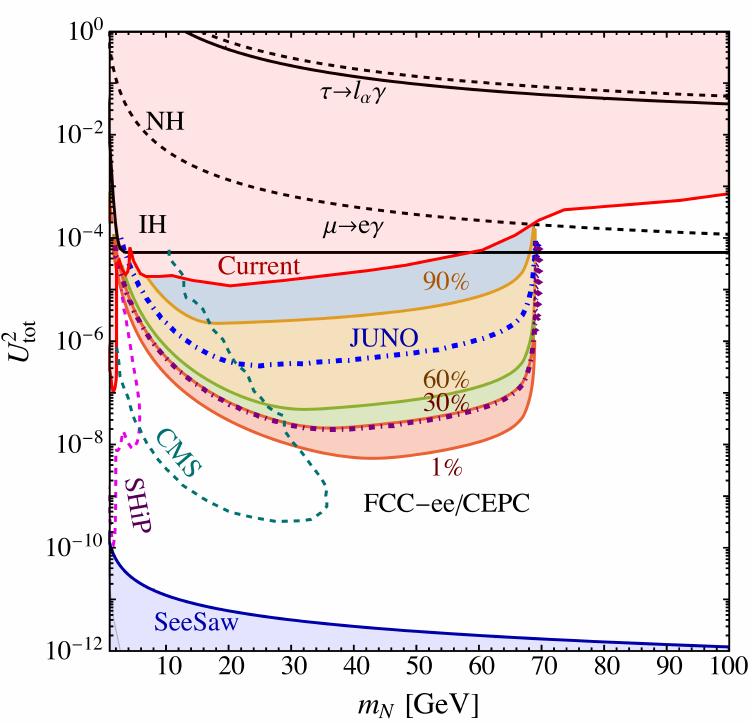}
    \caption{Fraction of the parameter space (in $\log_{10}$ scale) that allows unambiguous identification of the neutrino mass ordering (NMO) via    $e^+e^- \to N \nu\,(\bar{\nu})$, followed by $N \to \ell_\alpha j j$, at $\sqrt{s}=91.2$~GeV for CEPC/FCC-ee with  $\mathcal{L}\simeq 200~\mathrm{ab}^{-1}$.  Regions with $K>1\%~(30\%,\,60\%,\,90\%)$ and $K<30\%~(60\%,\,90\%)$ are shaded in red (green, orange, and cyan), respectively. 
    The red  area labeled ``Current'' is excluded by existing  constraints~\cite{Bolton:2019pcu}, while  dashed contours labeled ``SHiP'' and ``CMS'' indicate future sensitivities~\cite{Alekhin:2015byh,Drewes:2019fou}.
    The black line denotes the bound from $\ell_\alpha\to \ell_\beta\gamma$, for NH (dashed)  and IH (solid).
    The   ``Seesaw'' region reflects the cosmological bound $\sum_i m_{\nu_i} \simeq U_{\rm tot}^2\, m_N \lesssim 0.12~\mathrm{eV}$ with $i=1,2,3$ from Planck data~\cite{Planck:2018vyg}. 
    Blue/Purple dashed dotted curves are overlaid to show the interplay with JUNO's results for the indication of the physics beyond the two-HNL case with Scenario A/B. }
    \label{fig:sum}
\end{figure}

In Fig.~\ref{fig:sum} we show contours of the fraction $K$ in the
$(m_N,\,U_{\rm tot}^2)$ plane, together with existing bounds and future
sensitivities.
Cosmological observations from Planck constrain the sum of light neutrino
masses,
$\sum_i m_{\nu_i} \simeq U_{\rm tot}^2 m_N \lesssim 0.12~\mathrm{eV}$,
via the seesaw relation~\cite{Planck:2018vyg}.
Bounds from charged lepton radiative decays,
$\mathrm{Br}(\ell_\alpha\to\ell_\beta\gamma)\propto U_\alpha^2 U_\beta^2$,
are shown by the black curves for NH (solid) and IH (dashed), with details
given in the appendix.
The CEPC/FCC-ee sensitivity is kinematically suppressed as
$m_N \lesssim 70~\mathrm{GeV}$, 
since the two-body phase space suppresses  $\Gamma(Z\to N\nu)\propto (1-m_N^2/m_Z^2)^2$, leaving too few events to determine the neutrino mass ordering. 

Although CEPC/FCC-ee cannot directly probe the seesaw limit, they exhibit
strong discriminating power for the neutrino mass ordering:
$K\simeq1\%$ for $U_{\rm tot}^2\simeq4\times10^{-9}$, rising to
$K\simeq30\%~(60\%)$ for $U_{\rm tot}^2\simeq10^{-8}~(10^{-7})$.
Near-maximal discrimination, $K\simeq90\%$, as set by current oscillation parameter, can be reached for $U_{\rm tot}^2\gtrsim10^{-6}$.
This is about one order of magnitude below existing bounds (red solid line).
 \\

\noindent{\bf \textit{Interplay with JUNO's results}}
Since JUNO is expected to determine the neutrino mass ordering within the next five years, prior to the completion of future lepton colliders, it is interesting to explore the possible interplay with JUNO's results.

Here we consider two extreme scenarios to illustrate this interplay to show the best sensitivity of CEPC/FCC-ee to test two-HNL regime, as summarized in Table~\ref{tab:sce}. In scenario A, JUNO confirms the normal hierarchy (NH), while CEPC/FCC-ee have achieved the largest discrimination with only $e$ observation with  $R_{e} = (N_0+N_{B,e})/(N_0+\sum N_{B,\alpha})$. 
To establish this tension at the $2\sigma$ level, approximately $N_0 \simeq 3.7 \times 10^3$ events are required.

In scenario B, JUNO determines the IH. As an extreme case, we assume that CEPC/FCC-ee observes only the muon flavor, which would again be inconsistent with the JUNO result. Similarly, about $N_0 \simeq 8.2 \times 10^4$ events would be needed to reach the $2\sigma$ significance.
In Fig.~\ref{fig:sum}, the blue and purple dash-dotted curves indicate the regions where such deviations could signal physics beyond the minimal two-HNL framework for Scenario A/B.

\begin{table}
    \centering
    \small
    \begin{tabular}{|c|c|c|}
        \hline
        Scenario & JUNO & CEPC/FCC-ee  \\
        \hline
        A & NH Favor & $R_e=(N_0+N_{B,e})/(N_0+\sum N_{B,\alpha})$  \\
        \hline
        B & IH Favor & $R_\mu=(N_0+N_{B,\mu})/(N_0+\sum N_{B,\alpha})$  \\
        \hline
    \end{tabular}
    \caption{Scenarios for possible interplay between JUNO and CEPC/FCC-ee to test neutrino mass orderings at 2 $\sigma$.  }
    \label{tab:sce}
\end{table}

\noindent{\bf \textit{Conclusion}}
In this work, we have shown that the lepton-flavor structure of heavy neutral lepton interactions at future electron–positron colliders provides a sensitive probe of the neutrino mass ordering within the minimal Type-I seesaw framework. In scenarios with two nearly degenerate HNLs, the flavor composition of the heavy–light neutrino mixings is tightly correlated with the light-neutrino mass spectrum, leading to characteristic and experimentally testable signatures for normal and inverted orderings.

Focusing on future $Z$ factories such as CEPC and FCC-ee, we demonstrated that measurements of flavor-resolved HNL decays can probe mass-ordering–dependent parameter space over a wide range of HNL masses and mixings. While upcoming oscillation experiments are expected to determine the neutrino mass ordering through precision measurements at low energies, collider-based probes access fundamentally different information by testing the ultraviolet structure of neutrino mass generation. As such, collider measurements offer an independent and complementary validation of the neutrino mass spectrum within the seesaw paradigm.

A consistent picture emerging from both oscillation experiments and collider searches would provide strong evidence for minimal seesaw models with low-scale HNLs. Conversely, any discrepancy between low-energy determinations and collider-based flavor patterns would point to additional dynamics beyond the minimal framework, such as three-HNLs or non-standard interactions. In particular, once the neutrino mass ordering is determined by JUNO, we show two illustrative scenarios in which the flavor patterns measured at future lepton colliders could be inconsistent with the oscillation results, thereby signaling physics beyond the minimal two-HNL framework.
In this sense, collider probes of the HNL flavor structure remain valuable regardless of the experimental timeline for resolving the neutrino mass ordering, and they play a central role in testing the microscopic origin of neutrino masses.

\section*{Acknowledgments}

We thank F. F. Deppisch and Chayan
Majumdar for useful discussions. W. L. is supported by National Natural Science foundation of China (Grant No. 12205153).
J. S. was supported by IBS under the project code, IBS-R018-D1. The authors gratefully acknowledge the valuable discussions and insights provided by the members of the China Collaboration of Precision Testing and New Physics (CPTNP). 

\appendix
\section{Type-I seesaw framework}
\label{sec:model}

We consider a minimal Type-I seesaw with two heavy neutral leptons (HNLs)
$\nu_{RI}$ ($I=1,2$), sufficient to reproduce neutrino oscillation data.
The relevant Lagrangian is
\begin{equation}
\mathcal{L}
=
i\bar{\nu}_{RI}\slashed{\partial}\nu_{RI}
-
Y_{\alpha I}(\bar L_\alpha \cdot \tilde H)\nu_{RI}
-
\frac12 M_I \bar{\nu}_{RI}^c \nu_{RI},
\end{equation}
where $\alpha=e,\mu,\tau$ and $M_I$ is taken diagonal.
After electroweak symmetry breaking, $m_D=Yv/\sqrt2$, and the neutrino
mass matrix reads
\begin{equation}
\mathcal{L}_m^\nu
=
-\frac12
\begin{pmatrix}\bar\nu_L^c & \bar\nu_R\end{pmatrix}
\begin{pmatrix}
0 & m_D^T \\ m_D & m_R
\end{pmatrix}
\begin{pmatrix}\nu_L^c \\ \nu_R\end{pmatrix}.
\end{equation}

In the seesaw limit $m_D\ll m_R$, the light neutrino mass matrix is
$m_\nu\simeq -m_D^T m_R^{-1} m_D$.
The flavor eigenstates are related to mass eigenstates by
\begin{equation}
\nu_{L\alpha}
=
V^{\rm PMNS}_{\alpha i}\nu_i
+
\Theta_{\alpha I} N_I^c,
\qquad
\Theta_{\alpha I}\simeq \frac{vY_{\alpha I}}{M_I}.
\end{equation}

We employ the Casas--Ibarra parametrization~\cite{Casas:2001sr},
\begin{equation}
m_D
=
i\,V^{\rm PMNS}
\sqrt{m_\nu^{\rm diag}}\,
R\,
\sqrt{m_R^{\rm diag}},
\end{equation}
with $m_R^{\rm diag}={\rm diag}(M_1,M_2)$.
For normal (inverted) ordering, $m_1=0$ ($m_3=0$), and the complex
orthogonal matrix $R$ is
\begin{align}
R^{\rm NH} &=
\begin{pmatrix}
0 & 0\\
\cos\omega & \sin\omega\\
-\sin\omega & \cos\omega
\end{pmatrix},
&
R^{\rm IH} &=
\begin{pmatrix}
\cos\omega & \sin\omega\\
-\sin\omega & \cos\omega\\
0 & 0
\end{pmatrix},
\end{align}
where $\omega$ is complex and we define $x_\omega \equiv e^{\mathrm{Im}\,\omega}$.
Majorana phases are included through the PMNS matrix.

The charged-current interaction in the mass basis is
\begin{equation}
-\mathcal{L}_{\rm CC}
=
\frac{g}{\sqrt2}
W_\mu^+
\sum_{\ell}
\left(
\sum_{i=1}^3 V_{\ell i}^* \bar\nu_i
+
\sum_{I=1}^2 \Theta_{\ell I}^* \bar N_I^c
\right)
\gamma^\mu P_L \ell
+{\rm h.c.}
\end{equation}
For $M_1\simeq M_2$ and $\Theta_{\ell1}=\pm i\Theta_{\ell2}$, the two HNLs
form a pseudo-Dirac fermion with an approximate $U(1)$ symmetry.

HNLs decay via off-shell $W/Z$ bosons into leptonic and semileptonic final
states. The decay length in the laboratory frame is approximately~\cite{Drewes:2022rsk}
\begin{equation}\label{eq:lN}
L_N
\simeq
\frac{1.6}{U_{\rm tot}^2}
\left(\frac{m_N}{\rm GeV}\right)^{-6}
\left(1-\frac{m_N^2}{m_Z^2}\right)
{\rm cm}.
\end{equation}

Existing constraints on active--sterile mixing are summarized in
Ref.~\cite{Bolton:2019pcu}.
Mixing among different flavors induces radiative decays
$\ell_i\to\ell_j\gamma$ at one loop~\cite{Ibarra:2011xn,Dinh:2012bp}.
The strongest bound arises from MEG~II,
${\rm Br}(\mu\to e\gamma)<1.5\times10^{-13}$~\cite{MEGII:2025gzr},
with additional limits from $\tau\to\ell\gamma$ searches
\cite{BaBar:2009hkt,Belle:2021ysv}.
Future displaced-vertex searches at CMS~\cite{Drewes:2019fou} and the SHiP
beam-dump experiment~\cite{Alekhin:2015byh,SHiP:2018xqw} provide
complementary sensitivity, particularly in the long-lived HNL regime.

\section{Background at Collider}
\label{app:Bkg}
The irreducible four-fermion Standard Model background for CEPC and FCC-ee operating at the $Z$ pole, $e^+ e^- \to \ell \nu j j$, is generated at leading order using \textsc{MG5\_aMC@NLO}. The simulation includes contributions from both associated production via on-shell and off-shell $W$ bosons, as well as from $Z/\gamma^\ast$ production followed by radiation of an off-shell $W$ from one of the $Z$ decay legs. Other Standard Model backgrounds, including $ZZ$, $WZ$, $Z\nu\bar{\nu}$, top-quark, and multiboson processes, are neglected, as they either produce additional visible objects or require misidentification to mimic the exclusive one lepton and two jet final state. The resulting leading-order cross section for the background is found to be 0.72 fb per flavor.

\bibliographystyle{JHEP}
\bibliography{main}
\end{document}